\documentclass[aps,prb,twocolumn,showpacs,floatfix,amsmath,amssymb]{revtex4}
\usepackage{graphicx}
\usepackage{dcolumn}
\usepackage{bm}
\usepackage{color}

\begin{document}

\title{Spin-orbit coupling and the static polarizability of single-wall carbon nanotubes}
\author{Ginetom S. Diniz}
\email{ginetom@gmail.com} 
\affiliation{Department of Physics and Astronomy and Nanoscale
and Quantum Phenomena Institute, Ohio University, Athens, Ohio
45701-2979}
\altaffiliation{Current address: Institute of Physics, University of Bras\'{\i}lia, 70919-970, Bras\'{\i}lia-DF, Brazil}

\author{Sergio E. Ulloa}
\affiliation{Department of Physics and Astronomy and Nanoscale
and Quantum Phenomena Institute, Ohio University, Athens, Ohio
45701-2979}

\date{\today}

\begin{abstract}
We calculate the static longitudinal polarizability of single-wall carbon nanotubes in the long wavelength limit taking into account spin-orbit effects. We use a four-orbital orthogonal tight-binding formalism to describe the electronic states and the random phase approximation to calculate the dielectric function.  We study the role of both the Rashba as well as the intrinsic spin-orbit interactions on the longitudinal dielectric response, i.e. when the probing electric field is parallel to the nanotube axis. The spin-orbit interaction modifies the nanotube electronic band dispersions, which may especially result in a small gap opening in otherwise metallic tubes.  The bandgap size and state features, the result of competition between Rashba and intrinsic spin-orbit interactions, result in drastic changes in the longitudinal static polarizability of the system. We discuss results for different nanotube types, and the dependence on nanotube radius and spin-orbit couplings.
\end{abstract} 



\maketitle

\section{Introduction} 

Carbon nanotubes (CNTs) have been a subject of intense research since their discovery in 1992 and subsequent systematic synthesis by different growth techniques. \cite{Xavier} These quasi 1D-systems exhibit interesting physical properties that makes them ideal building blocks for nanodevices with a variety of proposed optoelectronic and sensor functionalities.  Among the different properties of CNTs, electronic transport is of special interest, as different structural details of the nanotubes (wrapping angle and diameter) result in qualitatively different electronic characteristics, including metallic and semiconducting behavior. \cite{Xavier}

The electronic properties of nanostructured devices are strongly influenced by the dielectric properties of the host material, and this is especially true of CNTs. Size and chirality dependence of the dielectric constant in CNTs have been studied previously, and shown to result in different behavior, depending on the electronic character, size and the external electric field configuration applied on the nanotube.\cite{Cohen} The zero frequency dielectric response of carbon nanotubes not only affects the charge carrier transport in nanoelectronics devices based on CNT, but can also be a tool to determine their interaction with adsorbed molecules and external fields. The different dielectric response of metallic and semiconducting CNTs has been shown to be an essential ingredient in sorting CNTs according to their electronic properties (as related to the different  polarizabilities), \cite{SORT} and to strongly affect the force microscope measurements on individual CNTs.\cite{Liwei1,Liwei2}

Interestingly, however, the effect of spin-orbit interactions (SOIs) has not been analyzed in the context of the polarizability of these structures. The existence of spin-orbit coupling in CNTs has been theoretically studied in previous works, \cite{Martino,Guinea,Rudner} and observed in transport experiments to have sizable effects, \cite{Flensberg} including breaking the four-fold degeneracy and electron-hole symmetry induced by the coupling of the electron orbital and spin degrees of freedom. \cite{Kuemmeth}  SOIs provide further fascinating opportunities
in the design of qubits and novel spintronic devices. \cite{QbitsCNT1,QbitsCNT2} In this work we explore the consequences of SOI on the polarizability of single walled CNTs.  The measure to which SOI effects may be tunable via applied  fields may provide in principle a useful manipulation scheme of the dielectric function of such nanoscale system.  This approach may also yield additional controls on the optimization of this material for electronic devices.

To this end, the aim of this paper is to present calculations of the longitudinal static polarizability of CNTs
in the long wavelength limit in the presence of SOI.
We consider SOI contributions known as the {\em intrinsic} spin-orbit (ISO), \cite{Kane} as well as the {\em Rashba} spin-orbit (RSO) interactions. \cite{Martino, Mahdi}
Our results are analyzed for different nanotube chiralities, and as function of radius and SOI parameters.
As we will describe in detail, ISO interactions induce the appearance or enhancement of an
energy gap in the spectrum near the Fermi level of undoped nanotubes (for metallic or semiconductor tubes, respectively).  This change is shown to be responsible for dramatic changes in the static polarizability, as the
intrinsic character of the CNT is modified.  In contrast, RSO interactions tend to reduce the gap and produce electron-hole asymmetries in the spectrum, which also affect the dielectric response, opening the possibility of a tunable dielectric function for nanoelectronic device implementation, \cite{Itkis} metallicity characterization, \cite{Krupke} and molecular detection.\cite{Heller}  Effects of SOIs on the transport properties have also been studied recently. \cite{DinizPRL}

In what follows, we describe the four-orbital tight-binding model used, and the effect of including spin-orbit couplings on the electronic band structure of different CNTs.  Similarly, we present the random phase approximation (RPA) formalism used to calculate the dielectric function, and explore the dependence of the polarizability on SOI parameters and  CNT characteristics.

\section{Theoretical Model}

\begin{figure}[!h]
\centering
\includegraphics[scale=0.20]{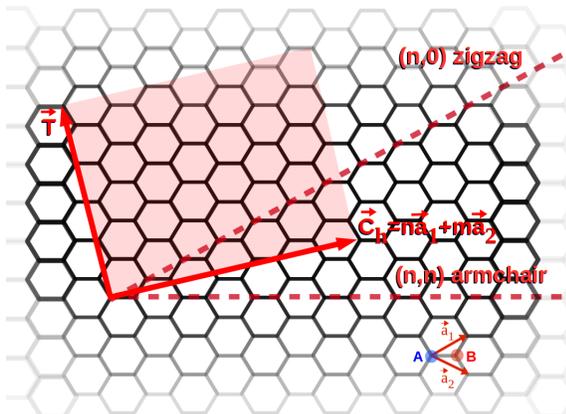}
\caption{(color online) Hexagonal lattice showing the two carbon atoms (red and blue circles, lower right)
of the unit cell of graphene.  Shaded region represents unit cell of the single-wall CNT, as
determined by the chiral vector $\vec{C}_{h}=n\vec{a}_{1}+m\vec{a}_{2}$, which defines the
edge of the rolled up CNT, and $\vec{T}$, which defines the axis of the tube.  In the zone-folding
approximation, CNTs have different electronic properties depending on the chirality indices $(n,m)$, as
indicated.}
\label{lattice}
\end{figure}

Single-wall carbon nanotubes can be seen as sections of a graphene sheet rolled up into a tube. Depending on the chiral vector that defines the edge of the rolled CNT, given by the indices $(n,m)$ as shown in Fig.\ \ref{lattice}, one can have metallic or semiconductor behavior. \cite{Saito}  Nanotubes with $(n,m=n)$ are called {\em armchair} for their edge profile and have always metallic behavior; those with
$(n,m=0)$ have {\em zigzag} edges, and can be both metallic or semiconducting.  As this classification follows the symmetries of flat graphene, it ignores effects of curvature and the concomitant orbital mixing that gives rise to sizable spin-orbit coupling of even the $\pi$-manifold. \cite{Guinea,LossPRL}
To model the carbon nanotube we use a four-orbital orthogonal tight-binding basis set,
considering nearest neighbors hopping and include SOI via two different terms in the Hamiltonian
\begin{equation}
H=H_{0}+ H_{ISO} + H_{RSO},
\label{Hamiltonian1}
\end{equation}
where $H_{0}$ is the graphene-like (no-SOI) Hamiltonian, with appropriate boundary conditions.
$H_{ISO}$ is the intrinsic spin-orbit term, and $H_{RSO}$ is the Rashba spin-orbit term, both present in a lattice system with broken mirror symmetry. The first term can be written as
\begin{equation}
H_{0}=\sum_{i,\sigma,\alpha}\epsilon_{i\sigma\alpha}c_{i\sigma\alpha}^{\dagger}c_{i\sigma\alpha} + \sum_{\langle i,j\rangle,\sigma,\alpha,\beta}t_{\alpha,\beta}^{ij}c_{i\sigma\alpha}^{\dagger}c_{j\sigma\beta} + H.c.,
\label{h0}
\end{equation}
where $c_{i\sigma\alpha}^{\dagger}$/$c_{i\sigma\alpha}$ creates/annihilates an electron,
$\epsilon_{i\sigma\alpha}$ is the on-site energy in different sites $i$ for different orbitals ($\alpha=2s,\, 2p_{x},\, 2p_{y}$, and $2p_{z}$) and spin, {$\sigma=\uparrow(\downarrow)$}. The parameter $t_{\alpha,\beta}^{ij}$ is the hopping parameter for different atomic orbitals, obeying the underlying graphene symmetries as well as the periodic boundary conditions intrinsic to the CNT. \cite{Saito} This Hamiltonian results in a $16\times16$ matrix, and its diagonalization gives spin up/down degenerate spectra as reported in previous work. \cite{Saito}

The spin-orbit interactions on the Hamiltonian are included only for the $\pi$-orbital subbands.
This approximation is excellent, as most of the contributions to the static dielectric function for undoped
CNTs comes from these subbands, given their close proximity to the Fermi level.
The intrinsic spin-orbit term is given by \cite{Kane}
\begin{equation}
H_{ISO}=i\lambda_{ISO}\sum_{\langle\langle i,j \rangle\rangle} \nu_{ij}c_{i}^{\dagger}s^{z}c_{j} + H.c.,
\label{hiso}
\end{equation}
where $\lambda_{ISO}$ is the intrinsic spin-orbit interaction, $s^{z}$ is the Pauli matrix and $\nu_{ij}=(2/\sqrt{3})|\hat{d}_{1}\times\hat{d}_{2} |=\pm1$, with $\hat{d}_{1}$ and $\hat{d}_{2}$ the two unit vectors defining the bonds that an electron chooses in going from site $i$ to $j$ (next-nearest-neighbors in the lattice). \cite{Kane} In $k$-space this ISO Hamiltonian can be written as,
\begin{equation}
H_{ISO}=\lambda_{ISO}\int d^{2}k\Psi^{\dagger}(k)M^{ISO}_\pi \Psi(k),
\end{equation}
where $M^{ISO}_\pi$ is a diagonal matrix in the basis of the $\pi$-orbitals,
$\Psi^T=(a_{k\uparrow}\: b_{k\uparrow}\: a_{k\downarrow}\: b_{k\downarrow})$, given by
\begin{eqnarray}
M^{ISO}_{\pi}=\left(\begin{array}{cccc}
f_{k} & 0 & 0 & 0\\
0     &-f_{k} & 0 & 0\\
0 & 0 &-f_{k} & 0\\
0 & 0 & 0 & f_{k}\\
\end{array}\right),
\end{eqnarray}
with $f_{k}=2\sin(\sqrt{3}a_{c}k_{y})-4\cos(3a_{c}k_{x}/2)$, $a_c=0.142$nm is the nearest neighbor
separation, and $k$'s are measured from the $\Gamma$ point.  

The third term in Eq.\ (\ref{Hamiltonian1}) is the Rashba interaction, which arises as one considers
extrinsic or built-in sources of electric field, typically assumed to be either in the radial direction of the nanotube
(due to curvature), or transverse to the CNT if arising from interaction with the substrate. \cite{LossPRL} 
One can write
\begin{equation}
H_{RSO}=i\lambda_{R}\sum_{\langle i,j \rangle} c_{i}^{\dagger}\left(\vec{u}_{ij}\cdot\vec{s}\right)c_{j} + H.c.,
\label{hsor}
\end{equation}
where $\vec{s}$ are the Pauli matrices. For a radial electric field (defined as the $\hat{r}$-direction),\cite{Martino}
one has $\vec{u}_{ij}=\hat{r}\times\vec{\delta}_{ij}$, where the vectors $\vec{\delta}_{ij}$ connect an atom A (or B) to
the three first-neighbors. The Rashba strength $\lambda_R$ is expected to be a few meV for graphene and related
carbon systems, \cite{Guinea} although recent experiments demonstrate that this value can be greatly increased to
about 200 meV when graphene is deposited on a Ni substrate. \cite{Dedkov}
Experiments in CNTs also point to sizable $\lambda_R$ values (of a few meV). \cite{Kuemmeth,Flensberg}
In $k$-space, the Rashba coupling can be written as
\begin{equation}
H_{R}=\lambda_{R}\int d^{2}k\Psi^{\dagger}(k)M^{R}_{\pi}\Psi(k),
\end{equation}
where, in the $\pi$-orbital basis one has
\begin{eqnarray}
M^{R}_{\pi}=\left(\begin{array}{cccc}
0 & 0 & 0 & iA +B\\
0 & 0 & -iA^{*}-B^{*} & 0\\
0 & iA-B & 0 & 0\\
-iA^{*}+B^{*} & 0 & 0 & 0\\
\end{array}\right),
\end{eqnarray}
with $A=-\sqrt{3}ie^{ia_{c}k_{x}/2}\sin(\sqrt{3}a_{c}k_{y}/2)$ and $B=e^{ia_{c}k_{x}}-e^{-ia_{c}k_{x}/2}\cos(\sqrt{3}a_{c}k_{y}/2)$.

The static polarizability is calculated using the random phase approximation, ignoring local field
effects. \cite{Cohen} The real part of the static dielectric function is given by
\begin{widetext}
\begin{equation}
\epsilon(\vec{q},\omega=0)=1+ \nu_{q}\sum_{p, n_{1},n_{2},s_1,s_2}\frac{\mid\langle p, n_{1}, s_1 \mid e^{-i\vec{q}\cdot\vec{r}} \mid s_2, n_{2}, p + q \rangle\mid^{2}}{E_{n_{1}s_1}(p)-E_{n_{2}s_2}(p+q)}
\, [f_{n_{2}s_2}(p + q)-f_{n_{1}s_1}(p)],
\label{e1}
\end{equation}
\end{widetext}
where $\nu_{q}=4\pi e^{2}/q^{2}\Omega$ is the 3D Fourier transform of the Coulomb interaction,
$\Omega$ is the volume, $n_{1}$ and $n_{2}$ refer to the quantized transverse momenta of the nanotube state
(also known as the subband index in CNTs), $p$ is the continuous momentum variable along the nanotube axis, 
$f_{n_{1}s_1}(p)$ is the Fermi function for the state $E_{n_1 s_1}(p)$, and $q$ is the longitudinal component of $\vec{q}$. Notice that as the states are mixed by
the spin-orbit interaction, they are not pure spin states and we sum over the two values of the `helicity' index $s_i=\pm$.   We now use Bloch expansions in the atomic orbitals of the carbon atoms,
\begin{eqnarray}
&\mid p, n_{1}, s_1\rangle=\sum_{\mu\sigma'} c_{\mu, \sigma'}^{n_{1},s_1}(p) \, \Phi_{1} \chi_{\sigma'} \nonumber \\
&\mid p+q, n_{2}, s_2\rangle=\sum_{\nu\sigma} c_{\nu, \sigma}^{n_{2},s_2}(p+q)\, \Phi_{2} \chi_{\sigma} ,
\end{eqnarray}
where the functions
\begin{eqnarray}
\Phi_{1} &=&\frac{1}{\sqrt{N}}\sum_{\vec{R^{\prime}},\tau_\mu}e^{i\vec{p}\cdot\left(\vec{R^{\prime}}+\tau_{\mu}\right)}\phi_{\mu}(\vec{r}-\tau_{\mu}-\vec{R}^{\prime})  \nonumber  \\
{\rm and}&& \nonumber \\
\Phi_{2}&=&\frac{1}{\sqrt{N}}\sum_{\vec{R},\tau_\nu}e^{i(\vec{p}+\vec{q})\cdot\left(\vec{R}+\tau_{\nu}\right)}\phi_{\nu}(\vec{r}-\tau_{\nu}-\vec{R}) ,
\end{eqnarray}
are written in terms of atomic orbitals $\mu$ (or $\nu$) at different lattice sites $\vec{R}$; $\tau_\mu =0, \vec{\delta}$, identifies the atomic basis in the unit cell, and $\chi_\sigma$ are the $\uparrow$ and $\downarrow$ spinors.
The matrix elements in Eq.\ (\ref{e1}) are proportional to the
orbital matrix elements%
\begin{eqnarray}
&&\langle \phi_{\mu}(\vec{r}-\tau_{\mu}-\vec{R}^{\prime})\mid e^{-i\vec{q}\cdot\vec{r}}\mid\phi_{\nu}(\vec{r}-\tau_{\nu}-\vec{R})\rangle \propto \nonumber \\
&&\langle \phi_{\mu}(\vec{r})\mid e^{-i\vec{q}\cdot\vec{r}}\mid\phi_{\nu}(\vec{r}-\vec{d})\rangle ,
\end{eqnarray}
with $\vec{d} = \vec{R} - \vec{R'} + \tau_\nu - \tau_\mu$. In the long wavelength limit, $e^{-i\vec{q}\cdot\vec{r}}\approx 1-i\vec{q}\cdot\vec{r}$, so that
\begin{eqnarray}
\lim_{\vec{q}\rightarrow 0} \langle \phi_{\mu}(\vec{r})\mid e^{-i\vec{q}\cdot\vec{r}}\mid\phi_{\nu}(\vec{r}-\vec{d})\rangle\approx\delta_{\mu \nu} \, \delta(\vec{d})-i\vec{q}\cdot\vec{R}_{\mu\nu} . 
\end{eqnarray}
The term $\vec{R}_{\mu\nu}(\vec{d})$ is the matrix element between the localized orbitals $\phi_{\mu}$ and $\phi_{\nu}$ centered at positions separated by $\vec{d}$.
The dipole matrix elements $R_{\mu\nu}(\vec{d})$ are vanishing small for non-zero $\vec{d}$, so that
the calculation can proceed by considering only the $\vec{d}=0$ on-site terms, as off-site elements
do not appreciably change the results. \cite{Cohen,LossPRL} The orbitals that contribute appreciably to the dipole matrix
element are those formed by the combination of the $2s$ and the $\{2p\}$ manifold.
The dipole matrix integral can be estimated by assuming hydrogenic
wave functions for the second shell of the carbon atoms and yield $R_{sp}=0.5$\AA. \cite{Cohen} 

We focus on the longitudinal dielectric response parallel to the CNT axis, with $\vec{q}\parallel \vec{p}$.  The 
transverse dielectric response for $\vec{q} \perp \vec{p}$, has small differences for metallic and semiconductor nanotubes, which by general  
arguments can be shown to depend quadratically on the radius.\cite{Cohen}  The transverse polarizabililty is then dominated by the finite size of the 
nanotube cross section and is anticipated to be less sensitive to SOI effects.

The static longitudinal polarizability in the long wavelength limit, ignoring local field corrections, is then given by
\begin{equation}
\label{unscreen}
\alpha = \frac{\Omega}{4\pi}  \lim_{\vec{q}_{\parallel}\rightarrow 0}
\left( \epsilon(\vec{q}_{\parallel},\omega=0) -1 \right) .
\end{equation}
The next section presents calculations of the polarizability for different CNTs and the sensitivity to SOI effects.

\section{Results}
The hopping parameters we have used for these calculations are based on the Tom\'anek-Louie parametrization for graphite; \cite{Tomanek} we measure the SOI parameters in terms of the hybridization constant
$t_{pp\pi}=-2.66$ eV.

\begin{figure}[!h]
\centering
\includegraphics[scale=0.9]{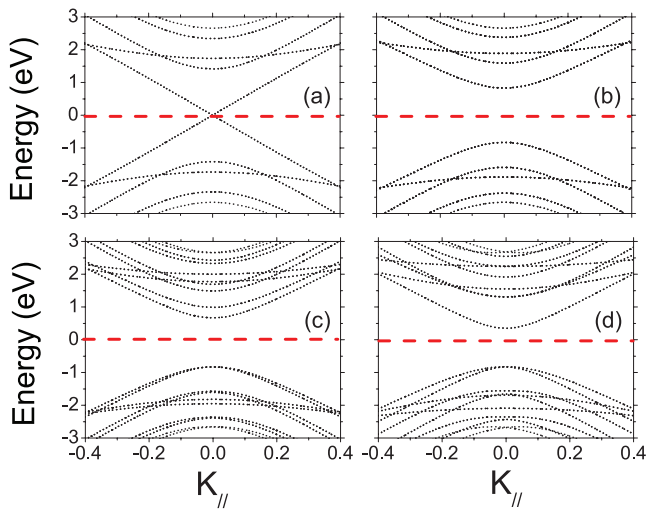}
\caption{Electronic structure for the zigzag nanotube (9,0) close to the Fermi level (dashed line at zero energy)
for different spin-orbit parameters.
(a) In the absence of SO interactions, the CNT displays linear dispersion near the Fermi energy
and metallic behavior. (b) With only ISO with $\lambda_{ISO}$=0.06$t_{pp\pi}\simeq 160$meV, the spectrum
exhibits a gap, but it does not lift the spin-double degeneracy nor the particle-hole symmetry. (c) Rashba and ISO contributions
with $\lambda_{ISO}$=0.06$t_{pp\pi}$ and $\lambda_{R}$=0.02$t_{pp\pi}$ break both electron-hole symmetry
and spin-degeneracy of the spectrum. (d) A larger Rashba coupling, $\lambda_{R}=\lambda_{ISO}=0.06t_{pp\pi}$, reduces the gap and enhances spin splitting.}
\label{90so}
\end{figure}

In Fig.\ \ref{90so}, we plot the electronic spectrum of the (9,0) zigzag nanotube for energies close to the
Fermi level (set at energy zero and indicated by the red dashed line).
In the absence of SO interactions this nanotube is metallic, as can be seen in panel \ref{90so}(a),
showing a spin-degenerate subband with linear dispersion near the Fermi level. Figure \ref{90so}(b) shows the
effect of a sizable ISO interaction ($\lambda_{ISO}=0.06t_{pp\pi}\simeq 160$meV), which is responsible
for a clear gap opening, but preserves particle-hole symmetry and spin-degeneracy of the spectrum.
In Fig.\ \ref{90so}(c) we have included the RSO interaction ($\lambda_{R}/t_{pp\pi}=0.02$), together with the
ISO term ($\lambda_{ISO}/t_{pp\pi}=0.06$). In addition to the gap opening, we see splitting of the various levels
due to the broken spin-degeneracy, resulting as well in a particle-hole asymmetric spectrum; this behavior has
been noted in previous work that includes both SOI interactions. \cite{Ralph}
In Fig.\ \ref{90so}(d) we consider both mechanisms, but with a stronger Rashba parameter,
$\lambda_{R}/t_{pp\pi}=0.06$. The larger $\lambda_R$ emphasizes the competition taking place between
both interactions: while ISO opens a gap near the Fermi level, the RSO term tends to reduce
it.  This behavior will have a clear signature on the static polarizability, as we will see below.
One can in fact obtain an analytical expression for this gap as function of both Rashba and ISO couplings,
so that the difference between the conduction and valence subbands gives an energy gap
$E_{g}=6\sqrt{3}\lambda_{ISO}-3\lambda_{R}$.
It is clear that the RSO competes with the ISO interaction, reducing the gap when both are present.

We should mention that for an originally semiconducting CNT the presence of SOI also affects the band
structure in similar ways, with competing Rashba and ISO interactions changing the value of the energy gap
and symmetries of the subbands.  These changes, however, are not nearly as qualitatively drastic as the strength of SOI is much smaller than typical original gaps. 

\begin{figure}[th]
\centering
\includegraphics[scale=0.6]{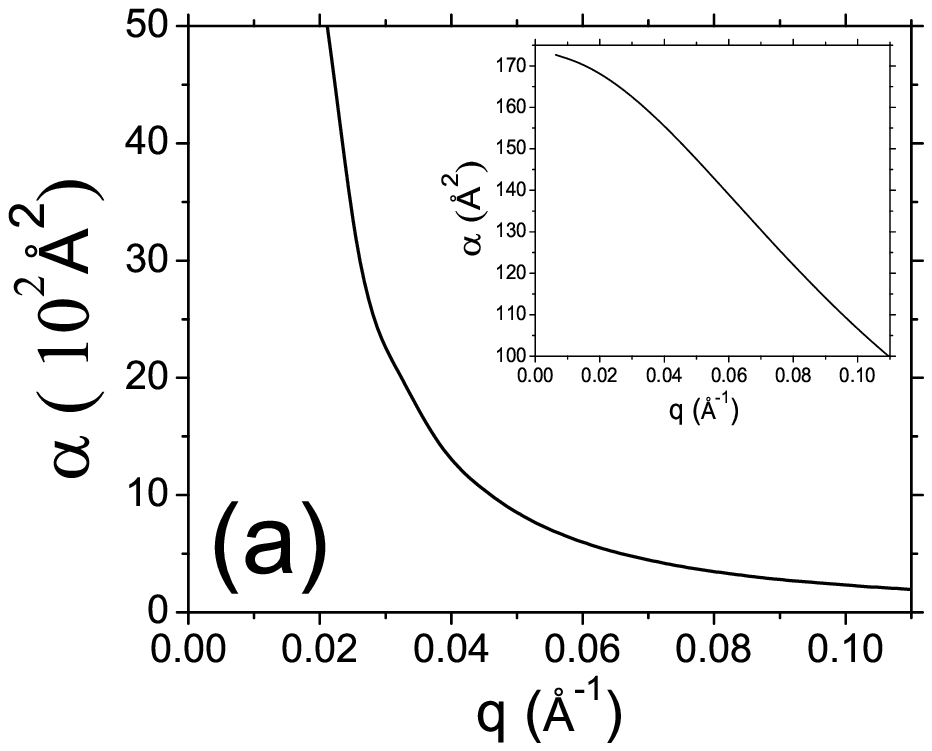}
\includegraphics[scale=0.6]{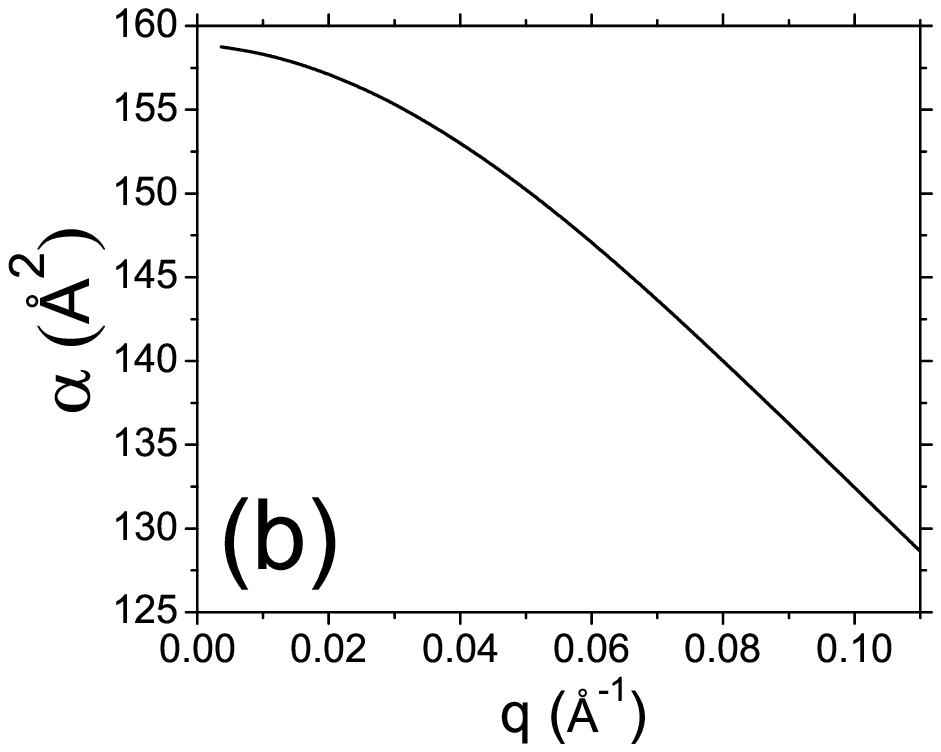}
\caption{Longitudinal static polarizability as function of wavenumber $q$ for different CNTs.
(a) For metallic armchair (5,5), $\alpha$ diverges as $q\rightarrow 0$, as one expects for perfect
screening.  Inset shows the same tube but with non-zero SOI present, $\lambda_{ISO}=0.02t_{pp\pi}$.  The
opening of a gap in the spectrum is reflected in the saturation of $\alpha (q=0)$. Similar behavior is
seen in (b) for the semiconductor zigzag tube (10,0).  Notice different vertical scales.}
\label{qlimit}
\end{figure}

Figure \ref{qlimit} shows the static polarizability as function of the reciprocal wavenumber near $q \simeq 0$,  
for different CNTs, with either metallic or semiconducting behavior.
For metallic nanotubes the static polarizability diverges as we approach the $q\rightarrow 0$ limit,  as seen in
Fig.\ \ref{qlimit}(a) for a (5,5) CNT.\@  This `perfect screening' behavior is expected from the static
Drude response for metallic systems $(\epsilon\left(q\right)\sim q^{-1})$ with non-zero density of states at the charge neutrality point. \cite{Katsnelson}
On the other hand, for semiconductor nanotubes such as the zigzag (10,0), the static polarizability in the long
wavelength limit is in fact finite, as can be seen in Fig.\ \ref{qlimit}(b).  This behavior is understandable
from the existence of a gapped spectrum in the system at the neutrality point.
In the presence of SOI, however, the generic opening of a gap in metallic tubes results in a finite
value for the polarizability, as shown explicitly in the inset of Fig.\ \ref{qlimit}(a)--notice the different vertical
scale--for the (5,5) CNT with $\lambda_{ISO}/t_{pp\pi}=0.02$.
This behavior is in general agreement with the expected linear response results for a gapped
system (and opposed to the Drude divergence for metallic systems).  We explore this in a more
quantitative fashion in what follows.

\begin{figure}[th]
\centering
\includegraphics[scale=0.8]{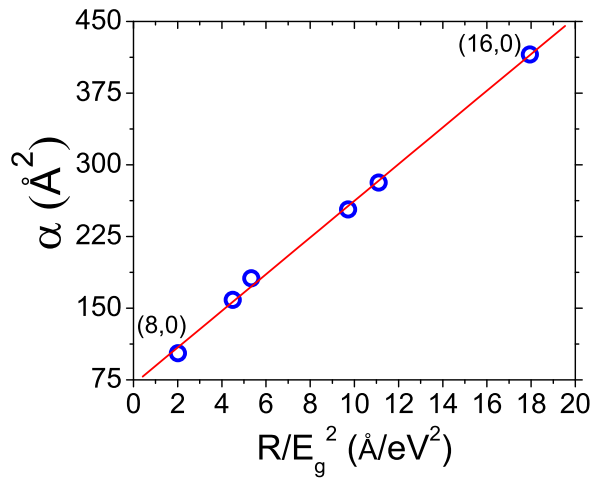}
\caption{(color online) Longitudinal static polarizability of zigzag (semiconducting) carbon nanotubes with
different $(n,0)$ indices at $q=0$.  
The circles represent the tubes $(n,0)$, with $n=8, 10, 11, 13, 14$, and 16. The polarizability
$\alpha$ depends nearly linearly on the variable $R/E_g^2$, as one expects from linear response
theory for a CNT with gap $E_g$ and radius $R$; the straight line is a simple fit to the data to guide the eye.}
\label{NoSograph}
\end{figure}

To set the framework for the discussion, Fig.\ \ref{NoSograph} shows our results for the longitudinal static polarizability
when {\em no} SOIs are considered; this figure shows different semiconductor zigzag nanotubes $(n,0)$, with $n=8, 10, 11, 13, 14$, and 16. 
Clearly, this picture does not consider armchair or metallic zigzag tubes, as they all show perfect screening in the long wavelength limit,
as discussed above.  There is a nearly linear dependence of the longitudinal static polarizability on
the variable $\sim R/E_{g}^2$, as expected from linear response theory. \cite{Cohen}
These results are also in agreement with explicit DFT calculations of the static polarizability, \cite{Kozinsky} and
with the general behavior seen in experiments. \cite{Fagan}

\begin{figure}[!h]
\centering
\includegraphics[scale=0.8]{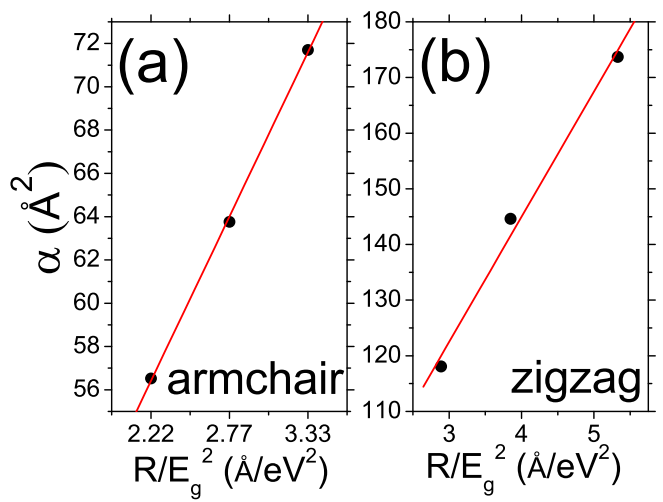}
\caption{(color online) Longitudinal static polarizability for different CNTs with ISO interactions, $\lambda_{ISO}=0.04t_{pp\pi}$ and $q=0$. 
(a) For armchair nanotubes (4,4), (5,5) and (6,6), $\lambda_{ISO}$ opens a gap, resulting in a finite $\alpha (q=0)$, which follows the linear dependence on $R/E_g^2$. (b) For zigzag nanotubes (9,0), (12,0) and (15,0), the linear dependence is also followed, if with a different slope.}
\label{tubeso}
\end{figure}

We now turn our attention to the effects of spin-orbit interaction on the static polarizability. As noticed before,
the inclusion of ISO interaction is responsible for a gap opening in the electronic spectra. Once
this energy gap is present, the perfect screening in the metallic nanotube is transformed into a {\em finite} polarizability, which
exhibits a similar linear dependence on the $R/E^2_g$ variable,
as can be seen in Fig.\ \ref{tubeso}(a) and (b). These graphs show the longitudinal static polarizability
of different CNTs with $\lambda_{ISO}=0.04t_{pp\pi}$. Panel (a) shows the armchair nanotubes (4,4), (5,5) and
(6,6) while (b) shows the zigzag tubes (9,0), (12,0) and (15,0). The important feature in these graphs is the finite
value of $\alpha$.  Notice that although the nature of the gap is due to ISO, the linear dependence on $R/E^2_g$ is still valid here, 
if only with  different slopes for the different families of CNTs.  

\begin{figure}[!h]
\centering
\includegraphics[scale=0.8]{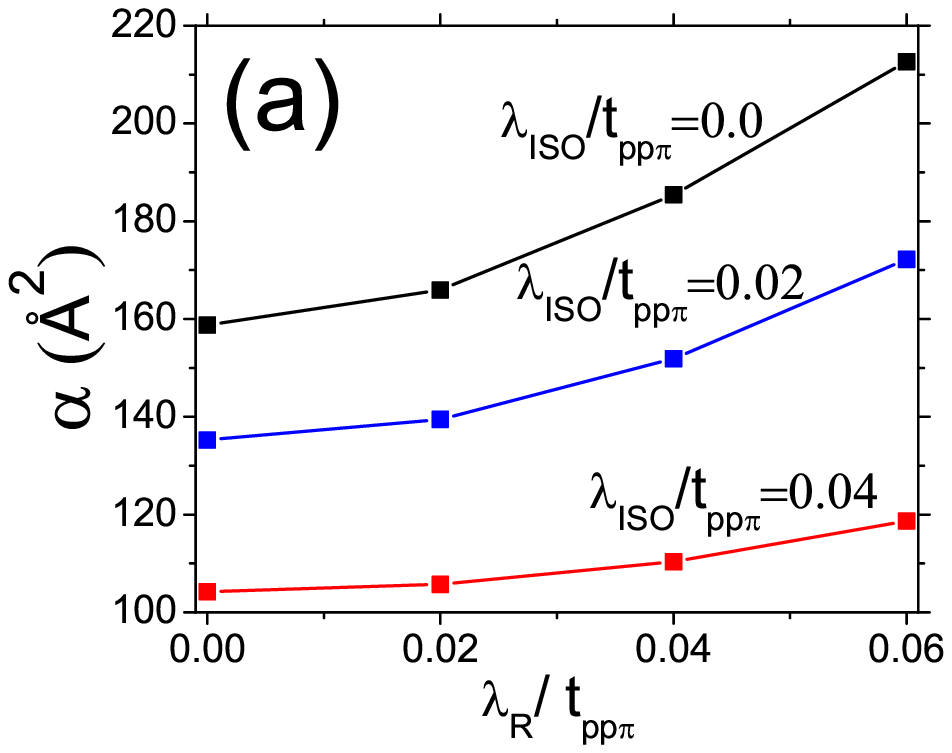}
\includegraphics[scale=0.8]{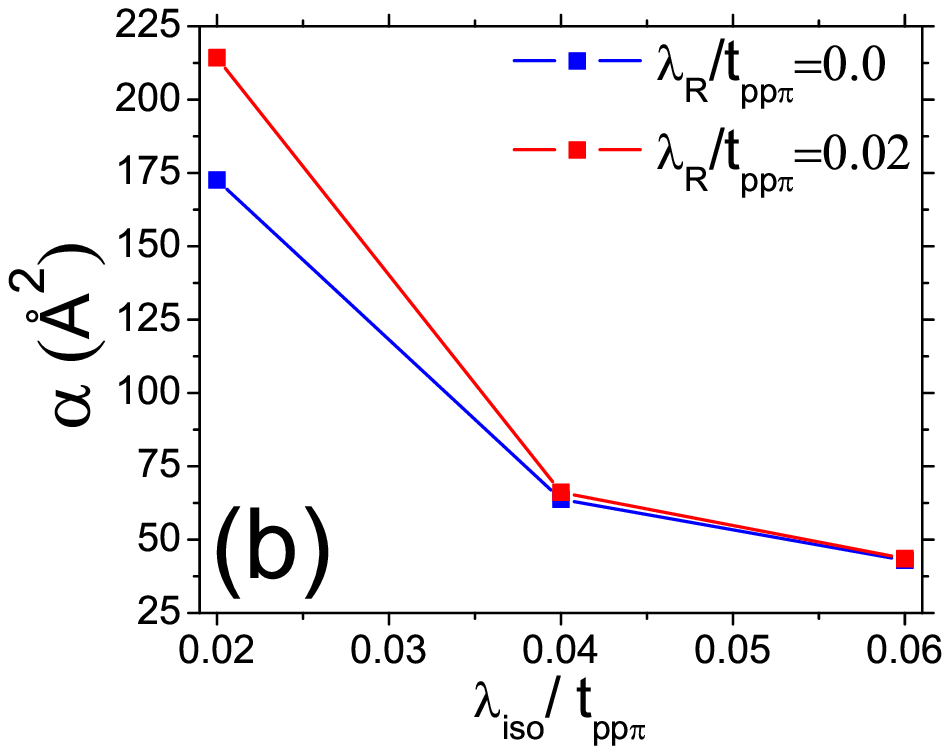}
\caption{(color online) Longitudinal static polarizability at $q=0$ for different CNTs and SOI. (a) $\alpha$ for semiconducting
zigzag (10,0) nanotube as a function of $\lambda_{R}$ for different values of $\lambda_{ISO}/t_{pp\pi}=0.0, 0.02$,
and 0.04.  Increasing $\lambda_R$ reduces the gap opened by $\lambda_{ISO}$, increasing $\alpha$.
(b) The armchair nanotube (5,5) as a function of $\lambda_{ISO}$ for different RSO parameters,
$\lambda_{R}/t_{pp\pi}=0.0$ and 0.02, shows the dominance of $\lambda_{ISO}$ in defining the gap
and polarizability.}
\label{tubeso2}
\end{figure}

We have also studied the behavior of the longitudinal static polarizability when both SOI contributions
are present.  As an example of a semiconducting CNT, we have calculated $\alpha$ for the (10,0) zigzag
nanotube. Figure \ref{tubeso2}(a) shows $\alpha$ as a function of $\lambda_{R}$ for different
$\lambda_{ISO}/t_{pp\pi}=0.0, 0.02$ and $0.04$. In the absence of SOI this nanotube is
semiconductor and the value of $\alpha$ is finite, as expected; once $\lambda_{ISO}$ is included, it produces
a reduction in $\alpha$, which is counterbalanced by an increasing RSO interaction,
as one would expect from the dependence of the fundamental electronic gap on SOI couplings.
We should stress that this behavior is in agreement with simple gap considerations, as the spin-mixed
character of the relevant states does not manifest itself in the polarization (which would relate to screening of charges).

To analyze $\alpha$ when both SOI contributions are present in an otherwise metallic system such as the
armchair (5,5) CNT, we show the longitudinal static polarizability vs.\ $\lambda_{ISO}$ for two different values
of $\lambda_{R}/t_{pp\pi}=0.0$ and 0.02. We notice that as $\lambda_{ISO}$ increases, $\alpha$ is
suppressed for a fixed $\lambda_R$, associated with the larger $E_g$ as $\lambda_{ISO}$ increases.
In contrast, one sees an enhancement of $\alpha$ when $\lambda_{R}$ increases.
This agrees with the expected behavior discussed in the previous figures.

\section{Conclusion}
We have studied the longitudinal static polarizability in the long wavelength limit for different
CNTs in the presence of different types of spin orbit interaction.
We used the random phase approximation to calculate the dielectric response, modeling the CNT
electronic spectra using a four-orbital orthogonal tight-binding formulation and the Tom\'anek-Louie parametrization.\cite{Tomanek}
Our results show that the metallic-semiconductor transition at neutrality that occurs when ISO interactions
are present, has a clear signature on the static polarizability, bringing the screening characterized by
 $\alpha$ to finite values for the case of metallic nanotubes or typically reducing it for semiconductor nanotubes.
As the RSO interaction produces a spin splitting/mixing of the bands, it would be interesting to see whether
this has consequences on the spin susceptibility or the screening of magnetic impurities.  

The drastic changes in polarizability as SOI are present suggest that perhaps one could use this effect to modulate
the dielectric response of the system via applied fields that enhance the Rashba coupling, for example.  It would be interesting
to investigate whether
this effect can be used to enhance the response to molecular adhesion selectivity or sensitivity.  

The results presented here may be useful in different applications: (i) in carbon nanotube separation processes, as the use of applied strong electric fields will intrinsically change the dielectric response; (ii) in molecular detectors, as a molecule attached to the walls of a carbon nanotube is responsible for strong local and likely radial electric fields, hence inducing a RSO-like interaction that can change the dielectric function; (iii)  in electric-field engineering of the dielectric response to provide desirable spin-dependent functionalities, including the possible polarization of current through CNTs. \cite{DinizPRL} 

\acknowledgements
We thank fruitful discussions with Liwei Chen, Mahdi Zarea, and Nancy Sandler. The authors acknowledge financial
support from Fulbright, CAPES (Brazil) and DMR/MWN (0710581 and 1108285) and PIRE NSF (0730257) grants.
SEU thanks the Dahlem Center for Complex Systems, FU Berlin, where parts of this work were completed, and the support of the
AvH Foundation.  


\end{document}